\documentclass[12pt,epsfig]{article}
\usepackage{graphicx}
\usepackage{amssymb,amsmath}
\begin{document}

\begin{titlepage}

\rightline{March 2010}

\vskip 2cm

\centerline{
\large \bf
Primordial $He'$ abundance implied by the mirror dark matter}
\vskip 0.3cm
\centerline{
\large \bf
interpretation of the DAMA/Libra signal}

\vskip 1.7cm

\centerline{P. Ciarcelluti \footnote{paolo.ciarcelluti@gmail.com}}
\vskip 0.5 cm
\centerline{
IFPA, D\'epartement AGO, Universit\'e de Li\`ege, 4000 Li\`ege, Belgium}

\vskip 1cm

\centerline{R. Foot \footnote{rfoot@unimelb.edu.au}}
\vskip 0.5 cm
\centerline{
School of Physics, University of Melbourne, 3010 Australia.}

\vskip 1cm

We compute the primordial mirror helium $He'$ mass fraction emerging from Big Bang nucleosynthesis in the mirror sector of particles in the presence of kinetic mixing between photons and mirror photons.
We explore the kinetic mixing parameter ($\epsilon$) values relevant for cosmology and which are also currently probed by the dark matter direct detection experiments.
In particular, we find that for $\epsilon \sim 10^{-9}$, as suggested by the DAMA/Libra and other experiments, a large $He'$ mass fraction ($Y_{He'} \approx 90\%$) is produced.
Such a large value of the primordial $He'$ mass fraction will have important implications for the mirror dark matter interpretation of the direct detection experiments, as well as for the study of mirror star formation and evolution. 

\end{titlepage}

During the last decade or so a huge experimental effort has been underway to directly detect dark matter and some interesting data are now available.
In particular the impressive annual modulation signal obtained by the DAMA/NaI experiment \cite{dama} has been confirmed with some precision in the DAMA/Libra experiment \cite{damalibra}, and represents exciting evidence for the discovery of dark matter. 
Mirror dark matter has emerged as one simple explanation \cite{mm1, mm2, mm3} for this and other experiments and thus deserves to be well studied in all its aspects.
Very briefly, mirror particles and forces are the hypothetical duplicate of the standard particles and forces with gravity usually assumed to be common to both sectors. 
The theory in the modern context of gauge theories was proposed in 1991 \cite{Foot:1991bp} although the basic idea is much older \cite{Lee:1956qn} (for a review, see Ref.~\cite{Foot:2004pa}).

The mirror dark matter paradigm implies a spectrum of dark matter particles with known masses, given by the masses of the stable nuclei and electrons ($H'$, $He'$, ..., $O'$, ...,  $Fe'$, $e'$...)\footnote{
As usual, mirror quantities are denoted with a prime ($'$).}.
The flat rotation curves observed in spiral galaxies can be explained if the galactic halos consist of an approximately spherically distributed gas of such mirror particles \cite{sph}.
Of course, in addition to this predominant spherical gas component, a more complex subcomponent is also possible and expected, consisting of mirror white dwarfs, stars with mixtures of ordinary and mirror matter, etc., which might ultimately be revealed in various ways, see e.g. Ref.~\cite{Sandin:2008}.

Besides gravity, mirror particles can couple to the ordinary ones via renormalizable photon-mirror photon kinetic mixing \cite{Foot:1991bp,foothe} with Lagrangian density
\begin{eqnarray}
{\cal L}_{mix} = {\epsilon \over 2}F^{\mu \nu} F'_{\mu \nu}
\end{eqnarray}
where $F^{\mu \nu} = \partial^{\mu} A^{\nu} - \partial^{\nu}A^{\mu}$ and $F'^{\mu \nu} = \partial^{\mu} A'^{\nu} - \partial^{\nu}A'^{\mu}$ are the field strength tensors for ordinary and mirror electromagnetism respectively.
This mixing enables mirror charged particles to couple to ordinary photons with charge $\epsilon qe$, where $q=-1$ for $e'$, $q=+1$ for $p'$, etc.
This interaction leads to important implications, not least is the possibility that mirror dark matter can be directly detected in experiments.
It turns out that the DAMA annual modulation signal \cite{dama,damalibra} can be fully explained \cite{mm1} via elastic scattering of the $\sim O'$  component of the halo, with: 
\begin{eqnarray}
\epsilon \sqrt{\frac{\xi_{O'}}{0.1}} \approx 10^{-9}
\end{eqnarray}
where $\xi_{A'} \equiv n_{A'}m_{A'}/(0.3 \ GeV/cm^3)$ is the halo mass fraction of the species $A'$.
Other, but more tentative evidence for mirror dark matter has emerged from the CDMS electron scattering data \cite{cdms}. 
It was shown \cite{mm2} that this data can be interpreted in terms of $e'$ scattering on electrons, and suggests
\begin{eqnarray}
\epsilon \approx 0.7\times 10^{-9} \ .
\end{eqnarray}
Finally, interpreting the two dark matter candidate events identified by the CDMSII/Ge analysis \cite{cdms2} as the expected $Fe'$ signal, yields the estimate \cite{mm3}
\begin{eqnarray}
\epsilon \sqrt{\frac{\xi_{Fe'}}{10^{-3}}} \approx 10^{-9}\ .
\end{eqnarray}
The compatibility of $\epsilon \sim 10^{-9}$ with cosmological constraints has been examined in Ref.~\cite{ciar}, where we showed that such mixing is consistent with constraints from ordinary Big Bang Nucleosynthesis (BBN) as well as the more stringent constraints from Cosmic Microwave Background (CMB) measurements and Large Scale Structure (LSS) considerations \cite{lss}.
Importantly, $\epsilon \sim 10^{-9}$ is also compatible with all experimental and astrophysical constraints, for a review see Ref.~\cite{reviewx}.

The mirror dark matter interpretation of the direct detection experiments depends on the mirror $He'$ mass fraction, $Y_{He'}$, and it is the subject of the present paper to compute it. 
The dependence comes about in the following way. 
Halo mirror particles form a self interacting plasma with a Maxwellian distribution:
\begin{eqnarray}
f_i (v) &=& e^{-\frac{1}{2} m_i v^2/T}  \nonumber \\  
        &=& e^{-v^2/v_0^2[i]} 
\label{8x}
\end{eqnarray}
where the index $i$  labels the particle type [$i=e', H', He', O', Fe'...$].
The dynamics of the mirror particle plasma has been investigated previously \cite{sph}, where it was found that the condition of hydrostatic equilibrium implied that the temperature of the plasma satisfied:
\begin{eqnarray}
T \simeq  {1 \over 2} \bar m v_{rot}^2 
\label{4}
\end{eqnarray}
where $\bar m = \sum n_i m_i/\sum n_i$ is the mean mass of the particles in the plasma, and $v_{rot} \approx 254$ km/s is the local rotational velocity for our galaxy \cite{rot}.
Assuming the plasma is completely ionized, a reasonable approximation since it turns out that the temperature of the plasma is $\approx \frac{1}{2}$ keV, then:
\begin{eqnarray}
{\bar m \over m_p} = {1 \over 2 - \frac{5}{4} Y_{He'}} \ .
\end{eqnarray}
Clearly, Eqs.(\ref{8x},\ref{4}) imply that the velocity dispersion of the particles in the mirror matter halo depends on the particular particle species and satisfies:
\begin{eqnarray}
v_0^2 [i] &=& {2T \over m_i} \nonumber \\
&=& v_{rot}^2 \frac{\overline{m}}{m_i} \nonumber \\
&=& v_{rot}^2 \frac{m_p}{m_i}\frac{1}{2 - \frac{5}{4}Y_{He'}} \ .
\end{eqnarray}
Thus, we see a dependence of the velocity dispersion of the halo dark matter particles on the parameter $Y_{He'}$. 
Consequently, event rates in dark matter experiments actually depend on $Y_{He'}$, and thus it is necessary and important to compute it.
It is also useful to know the primordial $He'$ mass fraction in order to study mirror star formation and evolution.
While the DAMA experiments turn out to be relatively insensitive to $Y_{He'}$, other experiments, such as electron scattering experiments, exhibit a greater sensitivity to $Y_{He'}$, and might ultimately be able to measure this parameter \cite{mm2}.

Previous works on BBN in the mirror sector \cite{previous} have parameterized the abundance of $He'$ in terms of some initial $T'/T$ value, without considering the effects of photon-mirror photon kinetic mixing.
Here we consider the implications of kinetic mixing for mirror BBN, and assume an effective initial condition of $T' \ll T$.
In the aforementioned papers it has already been emphasised that, compared with the ordinary matter sector, we expect a larger mirror helium mass fraction if $T' < T$, as currently required.
Essentially, this is because the expansion rate of the Universe is faster at earlier times, which implies that the freeze out temperature of mirror weak interactions will be higher than that in the ordinary sector.
Given the calculation of the $T'/T$ evolution of Ref.~\cite{ciar}, we can estimate the mirror helium mass fraction as a function of $\epsilon$ within the theory, which we now discuss.

We assume the initial condition $T'\ll T$. 
During the evolution of the early Universe the photon-mirror photon kinetic mixing populates and heats the mirror sector via the process $e^+ e^- \to e'^+ e'^-$.
The $e'^{\pm}$ will thermalize with $\gamma'$, however, because most of the $e'^{\pm}$ are produced in the low $T' \lesssim$ 5 MeV region, mirror weak interactions are too weak to significantly populate the $\nu'_{e,\mu,\tau}$.
Thus to a good approximation the radiation content of the mirror sector consists just of $e'^{\pm}$ and $\gamma'$.

From our earlier paper \cite{ciar}, we obtained an approximate analytical expression for $T'/T$ which is valid for $T' \gtrsim 1$ MeV and for $T \lesssim 100$ MeV, and is given by:
\begin{eqnarray}
{T' \over T} = \left(\frac{g}{g'}A\right)^{1/4} \left[ {1 \over T} 
             - {1 \over T_i}\right]^{1/4} \ .
\label{ana}
\end{eqnarray}
Here, we have assumed the initial condition $T' = 0$ at $T = T_i$ and
\begin{eqnarray}
A = \omega \times{27\zeta(3)^2 \alpha^2 \epsilon^2 M_{Pl} \over \pi^5 g\sqrt{g} }
\end{eqnarray}
where $\alpha$ is the fine structure constant, $M_{Pl}$ the Planck mass, $g$ ($g'$) the energetic effective degrees of freedom for ordinary (mirror) particles, and $\omega \approx 0.8$ takes into account the effect of various approximations (Maxwellian statistics instead of Fermi-Dirac, neglecting Pauli-blocking factors, etc.).
Assuming $T_i \gg 100$ MeV then Eq.(\ref{ana}) reduces to: 
\begin{eqnarray}
{T' \over T} \simeq \frac{0.25}{(T/{\rm MeV})^{1/4}} \sqrt{\frac{\epsilon}{10^{-9}}} \ .
\label{ana2}
\end{eqnarray}

In this theory the mirror sector starts with a temperature much lower than the temperature of the ordinary sector, and later the interactions induced by photon-mirror photon kinetic mixing increases only the temperature of mirror electrons, positrons and photons, since neutrinos are decoupled.
We may thus assume $T_{\nu'} \ll T'$, where $T' = T_{\gamma'} \simeq T_{e'}$, which is a reasonable approximation for the $\epsilon$ values of interest.
Thus, in this scenario the only reactions we need to consider to compute $Y_{He'}$ are 
\begin{eqnarray}
n' + e'^+ \to p' + \bar \nu' ~~~~~~ {\rm and} ~~~~~~ p' + e'^- \to n' + \nu' ~ .
\label{reactions}
\end{eqnarray}
We may neglect mirror neutron decay $n' \to p' + e'^- + \bar \nu'$, since the $n'$ lifetime is much longer than the available time for primordial mirror nucleosynthesis, which we estimate to happen in the first few seconds of the Universe.
The reaction rates of the processes (\ref{reactions}) can be adapted from the standard relations present in e.g. Ref.~\cite{Weinberg}, in which we can neglect the Pauli blocking effect on neutrinos because $T_{\nu'} \ll T'$:
\begin{eqnarray}
&&\lambda_{n'\rightarrow p'} = \lambda(n'+e'^+ \to p' + \bar \nu') 
  = B\int_0^\infty E_{\nu'}^2 p_{e'}^2 dp_{e'} [e^{E_{e'}/T'} + 1]^{-1} 
\nonumber \\
&&\lambda_{p'\rightarrow n'} = \lambda(p'+e'^- \to n' + \nu') 
  = B\int_{(Q^2-m_e^2)^{1/2}}^\infty E_{\nu'}^2 p_{e'}^2 dp_{e'} 
    [e^{E_{e'}/T'} + 1]^{-1} 
\label{1}
\end{eqnarray}
where 
\begin{eqnarray}
B = \frac{G_{\rm wk}^2 (1+3g_A^2) \cos^2\theta_C}{2 \pi^3 \hbar} \;,
\end{eqnarray}
$G_{\rm wk} = 1.16637 \times 10^{-5} \ {\rm GeV^{-2}}$ is the weak coupling constant, $g_{\rm A} = 1.257$ is the axial vector coupling of beta decay, measured from the rate of neutron decay, and $\theta_{\rm C}$ is the Cabibbo angle.
For $n'+e'^+ \to p' + \bar \nu', \ E_{\nu'} - E_{e'} = Q$, while for $p' + e'^- \to n' + \nu', \ E_{e'} - E_{\nu'} = Q$, where $Q \equiv m_n - m_p = 1.293$ MeV.
The extremals of integrals in Eqs.(\ref{1}) are fixed considering that integrations are taken over all allowed positive values of $p_{e'}$.

The differential equation for the ratio $X_{n'}$ of mirror neutrons to nucleons is:
\begin{eqnarray}
{dX_{n'} \over dt} = \lambda_{p'\rightarrow n'} (1 - X_{n'}) 
                   - \lambda_{n'\rightarrow p'} X_{n'} ~~.
\label{eq:Xn}
\end{eqnarray}
Note that $Y_{He'} \simeq 2X_{n'}$ since, as mentioned earlier, we can neglect $n'$ decay, and thus all available mirror neutrons go into forming $He'$.

\begin{figure}
  \centering
  \includegraphics[width=13cm,clip=]{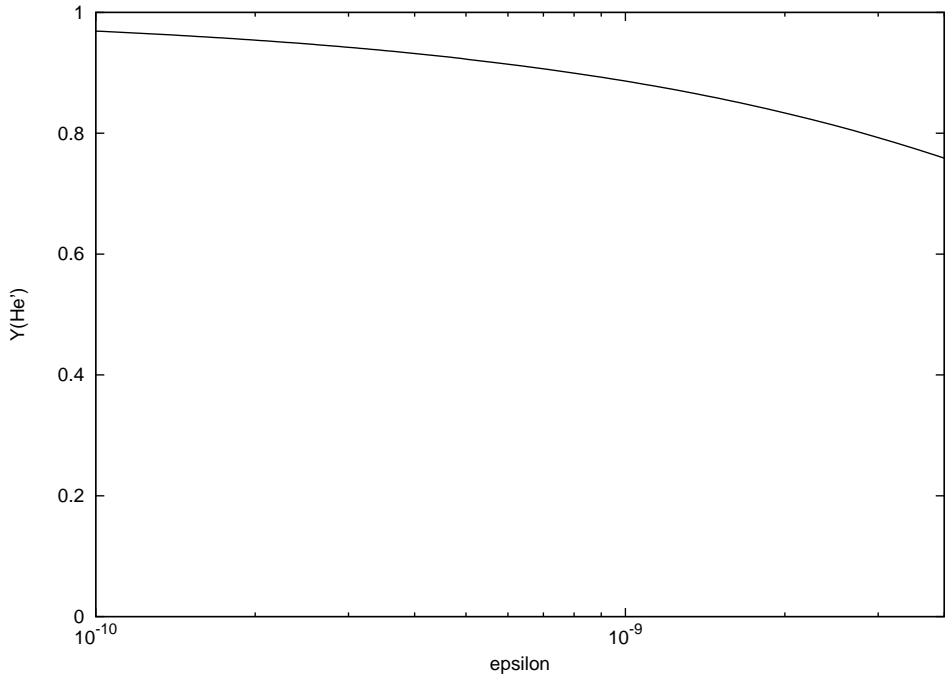}
  \caption{Mass fraction of primordial mirror helium ($Y_{He'}$) versus the strength of the photon-mirror photon kinetic mixing ($\epsilon$).}
  \label{fig:He-epsi}
\end{figure}

We have solved the above equations numerically, using the usual time-temperature relation for radiation dominated epoch
\begin{eqnarray}
t = 0.301 g^{-1/2} {M_{\rm Pl} \over T^2}
\end{eqnarray}
where $g$ takes into account only the degrees of freedom of ordinary particles, since the contribution of mirror particles is negligible given the initial condition $T' \ll T$.
We used the initial condition $X_{n'}(0)=0.5$ in Eq.(\ref{eq:Xn}), and followed the evolution until $X_{n'}$ reaches the asymptotic value.
Our results are shown in Figure \ref{fig:He-epsi}, where we plot the obtained mass fraction of mirror helium versus the strength of the photon-mirror photon kinetic mixing ($\epsilon$), for the parameter range of interest for cosmology.
As expected, we obtain high values of the primordial $He'$ mass fraction, with $ Y_{He'} \gtrsim 0.8$ for $\epsilon \lesssim 3 \times 10^{-9}$.
For the preferred value emerging from the analysis of the DAMA signal, $\epsilon \simeq 10^{-9}$, we obtain $ Y_{He'}\simeq 0.9$, which means that the dark matter is largely mirror helium dominated.

We can make a rough estimate of the primordial mass fraction of mirror elements of carbon mass and heavier. 
These are produced essentially via three-body interactions, the most important of which is the triple alpha process in the mirror sector: 
\begin{eqnarray}
^{4}He' + \:^{4}He' + \:^{4}He' \to \:^{12}C' + \gamma'. 
\end{eqnarray}
The rate for this process can be obtained from the rate for the corresponding process in the ordinary matter sector \cite{wagoner}, and is given by:
\begin{eqnarray}
{dX_{C'} \over dt} = 
  {1 \over 32} [Y_{He'}(t)]^3 \cdot 1.2\times 10^{-11} \rho_{b'}^2 \: T'^{-3} \,
  [\exp(-0.37 \: T'^{-1}) + 30.3 \exp(-2.4 \: T'^{-1})]
\end{eqnarray}
where $X_{C'}$ is the mass fraction of $C'$, $T'$ is in MeV units, $\rho_{b'}$ is the mirror baryon density in g/cm$^3$ and the rate is in sec$^{-1}$.
The $He'$ cannot be produced in significant proportion until $T' \lesssim 1$ MeV, which, from Eq.(\ref{ana2}), implies that $T \lesssim 7$ MeV (assuming $T_i > 100$ MeV).
Using $\rho_{b'} \approx 0.2 (T/{\rm MeV})^3 \ {\rm g/cm}^3$, then we estimate that the total mass fraction of $C'$ that can be produced is expected to be small, $X_{C'} < 10^{-8}$.

Anyway, this is just the primordial chemical abundance. 
Of course, light nuclei are expected to be processed into heavier nuclei by stellar nucleosynthesis on successive populations of mirror stars.
In this process of heavy element enrichment of the mirror dark matter interstellar medium a crucial role is played by the high fraction of $He'$ inside mirror stars.
In fact, accurate studies have shown \cite{Berezhiani:2005vv} that the stellar evolution is more rapid for a higher initial $He'$ content, and for $Y_{He'} \approx 0.9$ it can be orders of magnitude faster c.f. the standard case of $Y_{He} = 0.25$ in ordinary stars of the same masses.
This means that the enrichment of heavy mirror elements in the halo of the galaxy can be plausibly efficient enough to explain the relatively large $O'$ abundance suggested by the direct detection experiments \cite{mm1,mm2,mm3}.

In conclusion, we have shown that in the presence of a photon-mirror photon kinetic mixing of strength $\epsilon \approx 10^{-9}$, mirror dark matter is largely $He'$ dominated, with $Y_{He'} \approx 0.9$.
Merging this result with previous studies of mirror stellar evolution, we obtain a self consistent scenario for the mirror dark matter interpretation of the current experimental data. 
Our calculation of $Y_{He'}$ will play an important role in future precision tests of the mirror dark matter paradigm.

\vskip 1cm
\noindent

{\large Acknowledgements}
\vskip 0.2cm

\noindent
This work was supported by the Australian Research Council and by the Belgian Fund for Scientific Research (FNRS).

\end{document}